\documentclass[%
reprint,
superscriptaddress,
twocolumn,
 amsmath,amssymb,
 aps,
 prb,
]{revtex4-2}

\usepackage{graphicx}
\usepackage{dcolumn}
\usepackage{bm}
\usepackage{fixmath}
\usepackage{ulem}
\usepackage[usenames,dvipsnames]{color}

\newcommand{\twsetwo}{tWSe$_2$}

\begin{document}

\preprint{APS/123-QED}

\title{Hartree-Fock study of the moir\'e Hubbard model for twisted bilayer transition metal dichalcogenides}
\author{Jiawei Zang}
\affiliation{Department of Physics, Columbia University, 538 W 120th Street, New York, New York 10027, USA}
\author{Jie Wang}
\affiliation{Center for Computational Quantum Physics, Flatiron Institute, 162 5th Avenue, New York, NY 10010, USA}
\author{Jennifer Cano}
\affiliation{Center for Computational Quantum Physics, Flatiron Institute, 162 5th Avenue, New York, NY 10010, USA}
\affiliation{Department of Physics and Astronomy, Stony Brook University, Stony Brook, New York 11974, USA}
\author{Andrew J. Millis}
\affiliation{Department of Physics, Columbia University, 538 W 120th Street, New York, New York 10027, USA}
\affiliation{Center for Computational Quantum Physics, Flatiron Institute, 162 5th Avenue, New York, NY 10010, USA}
\date{\today}

\begin{abstract}
Twisted bilayer transition metal dichalcogenides have emerged as important model systems for the investigation of correlated electron physics because their interaction strength, carrier concentration, band structure, and inversion symmetry breaking are controllable by device fabrication, twist angle, and most importantly, gate voltage, which can be varied in situ. The low energy physics of some of these materials has been shown to be described by a ``moir\'e Hubbard model" generalized from the usual Hubbard model by the addition of strong, tunable spin orbit coupling and inversion symmetry breaking. In this work, we use a Hartree-Fock approximation to reach a comprehensive understanding of the moir\'e Hubbard model on the mean field level. We determine the magnetic and metal-insulator phase diagrams, and assess the effects of spin orbit coupling, inversion symmetry breaking, and the tunable van Hove singularity. We also consider the spin and orbital effects of applied magnetic fields. This work provides guidance for experiments and sets the stage for beyond mean-field calculations.
\end{abstract}

\maketitle


\section{\label{introduction} Introduction}
Twisted bilayer transition metal dichalcogenides (tTMD) have recently come to attention as important model systems for the investigation of basic issues in correlated electron physics \cite{wu2018hubbard,tang2020simulation,PhysRevResearch.2.033087}, due in part to the ability to tune electronic parameters over wide ranges by varying gate voltages without changing the device.  Experimental studies of twisted homobilayer WSe$_2$ (\twsetwo) \cite{Wang:2020us,ghiotto2021quantum,Li21} demonstrate interesting correlated electron behavior including continuous metal insulator transitions and ``bad metallic" and ``non Fermi liquid" transport. Unlike the delicate flat band in twisted bilayer graphene, which arises from phase cancellation of different hopping pathways and occurs only at specific ``magic angles" \cite{Bistritzer11}, the behavior of tTMD materials is controlled by correlation physics in relatively narrow bands, which can be achieved over a range of twist angles. Moreover, the monolayer components of tTMD materials have both a broken inversion symmetry and a strong spin orbit coupling, implying that the bands of tTMD materials also have these features. Consequences include a Dzyaloshinski-Moriya term in the spin Hamiltonian that describes strongly-coupled half filled bands and a gate voltage tunable shift in the energy position of the van Hove singularity. The spin orbit coupling also produces a relatively large ($9\sim13$ instead of $2$) $g$ factor \cite{forste2020exciton, lindlau2018role} which, with the narrow bandwidth and large unit cell, dramatically increases the sensitivity to applied magnetic fields. The ability to tune parameters over wide ranges in an experimentally accessible manner makes tTMD materials an important platform to  explore open problems in condensed matter physics and motivates theoretical studies. For example, recent  experimental studies of twisted homobilayer WSe$_2$ discovered a strange metal behavior near half filling and a metal-insulator transition that can be tuned continuously by varying gate voltages \cite{ghiotto2021quantum,Li21}.

Previous work \cite{wu2018hubbard,PhysRevResearch.2.033087,tang2020simulation,Wang:2020us,pan2020quantum,pan2020interactiondriven} has shown that the low energy physics of twisted homobilayers of TMD materials such as WSe$_2$ can be modelled as a variant of the triangular lattice Hubbard model, which we term the moir\'e Hubbard model. In this paper, we use Hartree-Fock calculations to achieve a comprehensive understanding of the moir\'e Hubbard model appropriate to tWSe$_2$  on the mean field level. We investigate the magnetic and metal-insulator phase diagram as a function of interaction, gate voltage and magnetic field, finding reentrant metal-insulator transitions driven by magnetic field and gate voltage at fixed carrier concentrations. We discuss the influence of the gate voltage dependent shift of the van Hove singularity on the phase diagram. Comparison of our work to experiments helps locate the experimental materials on the generic  Hubbard model phase diagram and opens up new directions for more accurate theoretical calculations.  

The rest of this paper is organized as follows. In Section~\ref{sec:Model} we present the model and parameters and describe their relation to the actual \twsetwo. In Section~\ref{sec:Method} we present the methods. In Section~\ref{sec:halffilling} we present the phase diagram as a function of gate voltage and magnetic field at half filling, and discuss the physical properties. In Section~\ref{sec:phase diagram all} we discuss possible magnetic ground states at general fillings. Section~\ref{sec:summary} is a summary and conclusion. Appendices present the details of our numerical methods. 

\section{Model \label{sec:Model}}
\begin{figure}[ht]
	\centering
	\includegraphics[width=0.9\columnwidth]{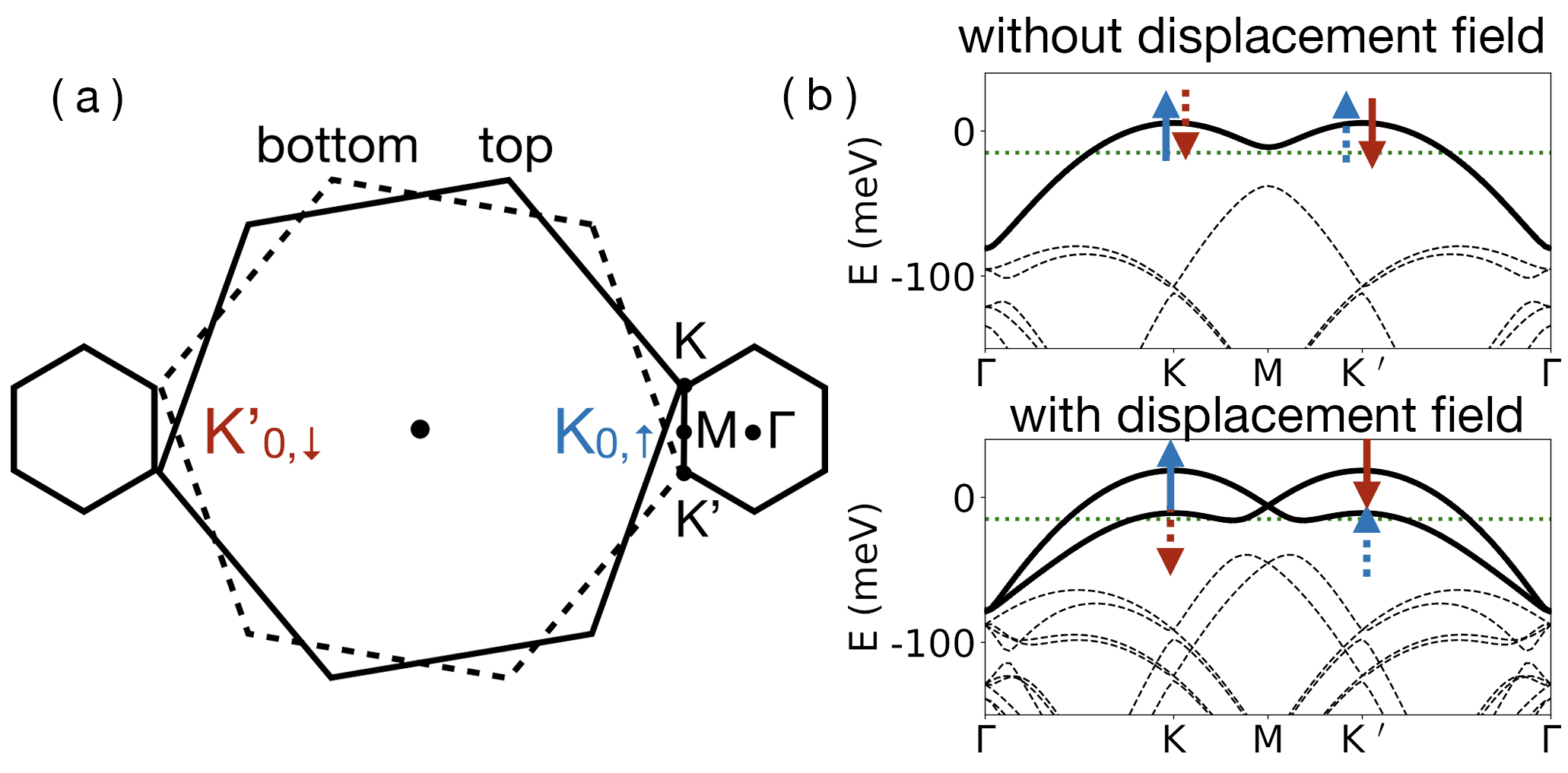}
	\caption{(a) Brillouin zones of the top (solid line) and bottom (dashed line) layer components of a twisted WSe$_2$ bilayer. Blue K$_{0,\uparrow}$ (red K$^\prime _{0,\downarrow}$) represents one valley with spin up (down) band at the valence band edges. Small hexagons indicate moir\'e Brillouin zones. (b) An illustrative band structure based on the continuum model of \twsetwo. We highlight the top most valence bands that can be matched to the Hubbard model. Blue solid (dashed) arrows represent the dominant spin of the top (bottom) layer at the $\vec{K}_0$ valley.
	Green dotted lines indicate the energy level of half filling of the topmost valence bands.}
	\label{brillouin}	
\end{figure}

In this paper, we focus on the twisted WSe$_2$ bilayer as an example of twisted homobilayer dichalcogenides.  In monolayer form, WSe$_2$ is a triangular lattice  semiconductor with inversion symmetry breaking and strong spin-orbit coupling (especially in the valence band). The top of the valence band occurs at the $\vec K_0$ and $\vec K'_0$ points of the  hexagonal Brillouin zone of the two dimensional monolayer (see Fig.~\ref{brillouin}). The strong spin-orbit coupling implies that the single-particle eigenstates have spin polarized perpendicular to the plane. Because of the strong inversion symmetry breaking, the highest-lying valence-band states dispersing downwards from the $\vec K_0$ point have spin up and the highest-lying valence band states dispersing downwards from the $\vec K'_0$ point have spin down, with a gap $\sim 0.4$ eV to the opposite spin states \cite{liu2013three}. 

Twisted WSe$_2$ is formed by stacking a second WSe$_2$ layer with a small commensurate twist angle. The resulting system is again a triangular lattice with a large ``moir\'e" unit cell and the corresponding ``moir\'e" Brillouin zone, with the $\vec K_0$ point of the top layer and the $\vec K'_0$ of the bottom layer mapping onto the moir\'e Brillouin zone $\vec K$ point, and conversely the $\vec K'_0$ point in the top layer and the $\vec K_0$ point in the bottom layer mapping onto the moir\'e Brillouin zone $\vec K'$ point (see Fig.~\ref{brillouin}(a)).

The highest valence bands of \twsetwo~may be understood \cite{wu2018hubbard,Wang:2020us} by taking the bands dispersing from the monolayer $\vec K_0$/$\vec K'_0$ points of each layer, back-folding them into the moir\'e Brillouin zone and then hybridizing them with a matrix element that is diagonal in moir\'e crystal momentum $\vec k$ and in spin. Details are given in Appendix \ref{app:band_structure}. The  strong spin-momentum locking of the individual layers and the momentum alignment, shown in Fig ~\ref{brillouin}(a), indicates that the spin up (down)  states near the moir\'e $\vec K$ point come predominantly from the top (bottom) layer. The broken inversion symmetry of the individual layers leads to inversion symmetry breaking in the the moir\'e system, which however retains a $C_3$ three-fold rotation symmetry and, if the two layers are identical, a $C_{2x}$ two-fold rotation symmetry that swaps the two layers. The combination of $C_{2x}$ and time reversal symmetry leads to a band degeneracy along high symmetry lines from $ \vec{\Gamma}$ to  $\vec K/\vec K^\prime$ and $\vec K/\vec K^\prime$ to  $\vec M/\vec M^\prime$, as seen in Fig.~\ref{brillouin}(b), upper panel. Application of a transverse ``displacement field'' (interlayer potential difference tuned by the top and bottom gate voltages, conventionally denoted as $D$) breaks the $C_{2x}$ symmetry between planes, lifting the degeneracy along these high symmetry directions and changing the band structure significantly,  as shown in Fig.~\ref{brillouin}(b), lower panel.

Even for zero displacement field, $D=0$, the moir\'e single particle eigenstates at a general wavevector $\vec k$ are non-degenerate \cite{Senthil_TBG_Sym,2020arXiv201003589W}.  However, for small twist angle (many atoms in the moir\'e unit cell) and weak interlayer hybridization we may restrict our attention to monolayer states very near the single layer $\vec K_0 /\vec K_0 ^\prime$ points, so that the  single layer valence band may be approximated as a parabola $\varepsilon_{\vec k}=-(\vec{k}-\vec{K}_0)^2/2m^{\ast}$ (``continuum model''). In this approximation the moir\'e system has an emergent  inversion symmetry ($E_\sigma(\vec{k})=E_\sigma(-\vec{k})$) if the two individual layers are identical, so combining with time reversal symmetry, at $D=0$ the bands at any $\vec k$ point would be spin degenerate. This degeneracy is broken by terms of order $|\vec{k}-\vec{K}_0|^3$ in the monolayer band structure \cite{Korm_nyos_2015}. These cubic terms  have effects that are small by a factor of the order of the inverse of the number of atoms in the moir\'e unit cell. We neglect these small terms here, so that the model we study is fully inversion symmetric at $D=0$ with inversion symmetry broken by the displacement field. 

The result of these considerations is that the one-electron properties of the top of the valence band of \twsetwo~can be described by a tight binding model with hopping $c_{i, \sigma}^{\dagger} t^{i,j}_{\sigma}  c_{j, \sigma}$, where $t^{i,j}_{\sigma}=|t|e^{i\sigma\phi_{ij}}$. $\sigma$ indicates spin and also valley due to the spin-valley locking, and the phase $\phi$ parametrizes the inversion symmetry breaking arising from a non-zero displacement field.
Ref.~[\onlinecite{Wang:2020us}] shows that we need only to retain the nearest neighbor hopping, with a second neighbor term  $\sim 20\%$  of the first neighbor term. Our convention for $\phi_{ij}$ for nearest neighbor hopping is shown in Fig.~\ref{fig:phi}(a). At zero displacement field $t^{i,j}$ may be taken to be independent of $\sigma$ (up to terms of order of the inverse of the number of atoms in the moir\'e unit cell, which we neglect); as the displacement field is increased, the spin dependence of $t$ becomes more pronounced and the magnitude of $t$ changes. Previous work also indicates that the important interaction effects come from an on-site repulsion, so the twisted bilayer material is governed by the generalized ``moir\'e" Hubbard Hamiltonian with only nearest neighbor hopping \cite{wu2018hubbard,PhysRevResearch.2.033087}:
\begin{align}
	H&=-\sum_{\substack{\vec{k},\vec{a}_{m},\\ \sigma=\pm}}2|t|\cos (\vec{k} \cdot \vec{a}_{m}+\sigma \phi)c^{\dagger}_{\vec{k},\sigma}c_{\vec{k},\sigma}+U \sum_{i} n_{i \uparrow} n_{i \downarrow},
	\label{eq:H}
\end{align}
where $\vec{a}_{m=1,2,3}$ are the lattice vectors, $\vec{a}_1=a_M(1,0),~\vec{a}_2=a_M(-\frac{1}{2},\frac{\sqrt{3}}{2}),~\vec{a}_3=a_M(-\frac{1}{2},-\frac{\sqrt{3}}{2})$, and $a_M$ is the moir\'e cell lattice constant.  From previous DFT calculations \cite{Wang:2020us}, physically achievable values of $D$ correspond to changing $\phi$ over the range $0\lesssim\phi\lesssim \pm\frac{\pi}{3}$, and increasing the magnitude of $|t|$ from $t_0$ (hopping amplitude at zero displacement field) to $\sim1.3 t_0$. In this work, we set $|t|=1$ as the unit of energy scale, and thereby $U$ represents the ratio of on-site interaction and the hopping amplitude $|t|$. Due to the spin-valley locking, the sum over spins in Eq.~(\ref{eq:H}) is also a sum over both valleys; consequently, in coupling spins, the $U$ term also couples the two valleys.

In this model changing $\phi \leftrightarrow -\phi$ interchanges spin up and spin down, and a particle-hole transformation $|t|\rightarrow -|t|$ corresponds to $\phi\rightarrow \phi-\pi$, so that  the physics  can be entirely reconstructed from the physics of $0<\phi<\pi/2$ by a combination of particle-hole transformation and spin inversion.

The nearest neighbor hopping model has additional symmetries which may be understood by considering the spin-dependent phase factor in the hopping $|t|e^{i\sigma \phi}$ as  either a spin dependent Peierls phase factor arising from a spin-dependent gauge field or a position-dependent spin rotation.  Taking the first point of view we observe that a DM field characterized by an angle $\phi$ corresponds to a system in a spatially varying magnetic field producing a flux of $\pm 3\phi$ through each triangular plaquette.  The flux is opposite for the two spin directions and changes sign between the two sublattices of the dual lattice formed by the centers of the triangular plaquettes.  The form $|t|e^{i\sigma\phi}$ is a gauge choice consistent with this flux. Changing $\phi\rightarrow\phi+2\pi/3$ corresponds to introducing a flux of $\pm 2\pi$ per plaquette which which does not change the spectrum (although as discussed below it does change the wavefunction). Changing $\phi\rightarrow\phi+\pi/3$ corresponds to introducing a flux of $\pm \pi\equiv 3\pi$ per plaquette. A phase change of $\pi$ on each link is equivalent to a particle-hole transformation, so the spectrum at $n,\phi$ is the same as the spectrum at $2-n,\phi\pm \pi/3$.

While the spectrum is invariant under certain changes in $\phi$, the wave function (and therefore the magnetic ordering pattern) will change. To see this, note that a space-dependent rotation of the electron spin by an angle $2\gamma_i$ about the z axis is implemented by the matrix $R_i=e^{-i\gamma_i\sigma_z}$ and leads to the change $t_{ij}\rightarrow t_{ij}e^{i(\gamma_i-\gamma_j)\sigma_z}$. Thus, the DM field can be thought of as a space-dependent spin rotation.  


Twisted WSe$_2$ has a large $g$ factor and a large moir\'e unit cell compared to usual untwisted materials. Thus, it is interesting to consider the spin and orbital effects of the magnetic field perpendicular to the lattice. The strong spin orbital coupling characteristic of monolayer WSe$_2$ implies a Zeeman interaction term $H_1=-g \mu_BBS_z$ with $g\sim9-13$ \cite{forste2020exciton, lindlau2018role}. The consequences of the Zeeman interaction will be discussed in Section \ref{sec:halffilling}. In addition, $t_{ij}$ will pick up an additional phase $\psi_{AB}^{i,j} \approx\left[(e / \hbar) \int_{i}^{j} \vec{A} \cdot d \vec{r}\right]$ where $\vec{A}$ is the vector potential, due to the Aharonov-Bohm effect.  The phase can be thought of as  proportional to the flux through a closed loop of a triangular plaquette of a moir\'e unit cell. The area of the ``moir\'e'' unit cell is estimated to be $S$ $\approx\frac{a^{2} \sqrt{3}}{4(1-\cos \theta)}\propto 1/\theta^2$. Table~\ref{table} shows the estimated phase $\psi_{AB}$ per triangular plaquette of a unit cell in \twsetwo~with monolayer lattice constant $a$ = 0.328~nm.  

We see that for small twist angle, achievable fields can produce a flux per unit cell of order 1. But for the twist angle $>3^\circ$ used in recent experiments, the orbital effects are much smaller. 
Thus, we do not consider these effects any further.

\begin{table}[ht]
\caption{\label{table}%
Estimation of $\psi_{AB}$ due to the magnetic field in \twsetwo~at different twist angles.}
\begin{ruledtabular}
\begin{tabular}{ccccc}
B (T) & $1^{\circ}$ & $2^{\circ}$ & $3^{\circ}$& $>3^{\circ}$\\ \hline
5 & 0.37$\pi$ & 0.09$\pi$ & $0.04\pi$ & $<0.04\pi$\\ 
10 & 0.74$\pi$ & 0.18$\pi$& $0.08\pi$ & $<0.08\pi$\\
\end{tabular}
\end{ruledtabular}
\end{table}

\section{Method\label{sec:Method}}
We solve the model in the Hartree-Fock approximation, focusing on the effects of a non-zero displacement field.  For orientation it is useful to summarize previous considerations of the half filled large $U$ limit, in which the low energy physics is described by a Heisenberg model with an interesting dependence on the displacement field \cite{PhysRevResearch.2.033087}. The Hamiltonian in this limit is given by:
\begin{eqnarray}
    H &=& \sum_{\langle ij \rangle}JS_i^zS_j^z+J\cos2\phi\left(S_i^xS_j^x+S_i^yS_j^y\right)\nonumber\\
    &+& J\sin2\phi~ \hat{ e}_z\cdot ( \vec{S}_i\times \vec{S}_j ).
    \label{Heisenberg}
\end{eqnarray}
Here $\vec{S}$ is the vector of $S=\frac{1}{2}$ Pauli matrices and $\langle ij \rangle$ denotes nearest neighbors. 

\begin{figure}[t]
  \includegraphics[width=0.9\columnwidth]{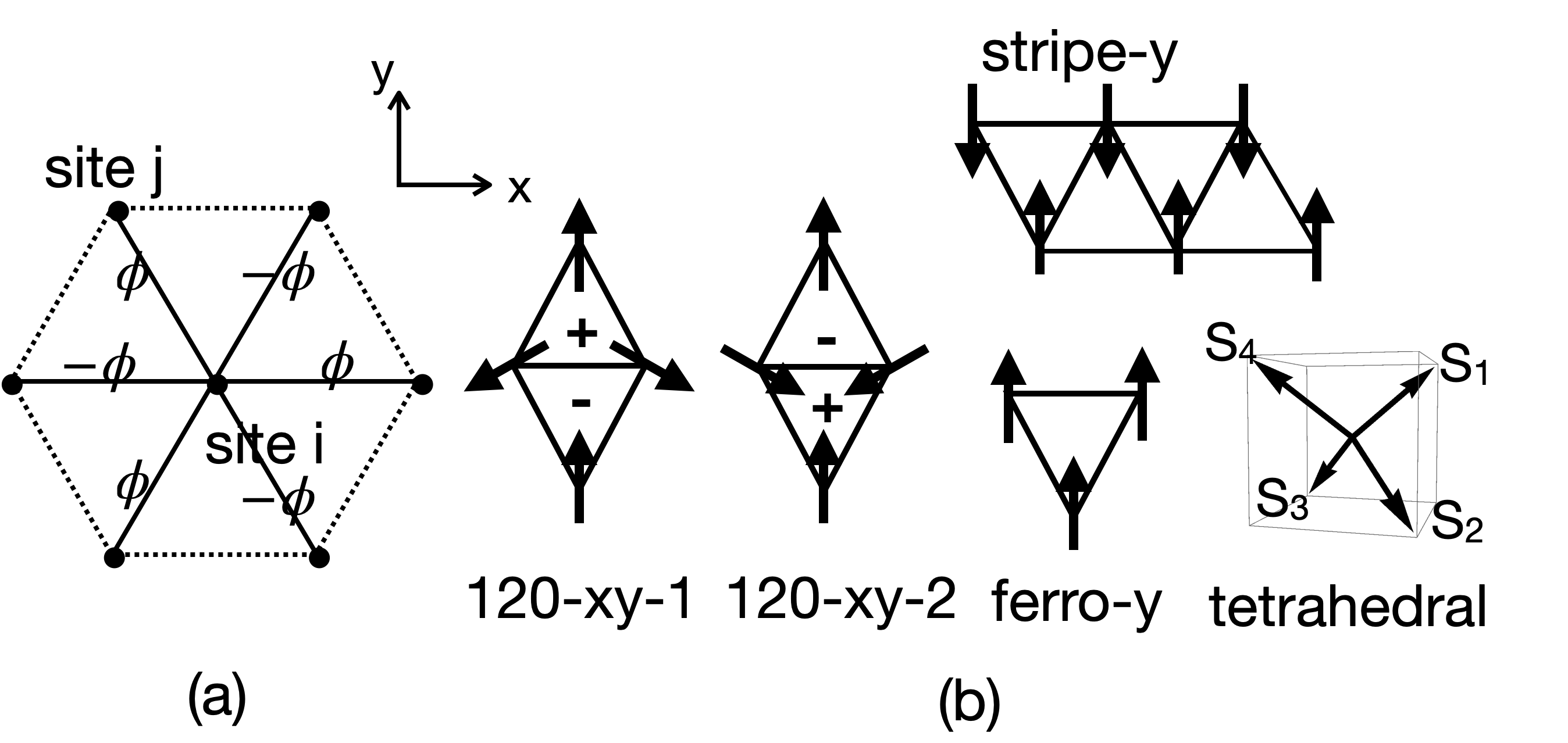}
  \caption{(a) Sketch of the phase $\phi_{i,j}$ between a given site i and its neighbor site j on a triangular lattice, which is chosen based on symmetry. 
  (b) Possible  magnetic ordering patterns. For the tetrahedral order, the magnetic order on site i is defined as $\frac{S}{\sqrt{3}}(\cos \vec{Q}_0\cdot \vec{R}_i,\cos \vec{Q}_1\cdot \vec{R}_i,\cos \vec{Q}_{-1}\cdot \vec{R}_i)$, with
$\vec{Q}_0=(2\pi,0),\vec{Q}_{\pm1}=(-\pi,\pm\sqrt{3}\pi)$. The magnetic moment directions on the lattice are specified by the corresponding arrows shown in the tetrahedral sketch. For $120^{\circ}$ order, there are two chiralities. "+" and "-" are defined by the sign of  $\kappa=\frac{2}{3 \sqrt{3}}(\vec{S}_1 \times \vec{S}_2+\vec{S}_2 \times\vec{S}_3+\vec{S}_3 \times \vec{S}_1 )\cdot\hat{\vec e}_z$. We only draw the basic patterns; others can be generated by applying appropriate symmetry operations.}
  \label{fig:phi}
\end{figure}

Possible ordering patterns are shown in Fig.~\ref{fig:phi}(b). At $\phi=0$ the Heisenberg model exhibits $120^\circ$ order \cite{krishnamurthy1990mott,Jayaprakash_1991}. An alternative striped state is found to be slightly higher in energy, as is a tetrahedral state with a non vanishing $\vec{S}_{i} \cdot (\vec{S}_{j} \times \vec{S}_{k})$ on each triangular plaquette \cite{martin2008itinerant}. As discussed in Ref.~\cite{martin2008itinerant,pasrija2016noncollinear} and below, the tetrahedral state is favored at electron density $n$=1.5 and weak coupling. At $\phi=0$ the magnetic states have a high degree of ground state degeneracy. For the 120$^\circ$  state  the spins lie in a plane and there is a family of ground states characterized by $O(3)$ rotations of the vector normal to the plane. In addition the ground states are degenerate under a uniform rotation of all spins about the axis normal to the plane. Finally,  the ground state is characterized by a staggered chirality (sense of rotation of spins about a triangle) to which corresponds a $Z_2$ degeneracy.  For $\phi\neq 0,\pi$ the situation is different. The Dzyaloshinski-Moriya (DM) term $\hat{ e}_z\cdot (\vec{S}_i\times \vec{S}_j)$ breaks the $O(3)$ invariance \cite{dm1,dm2}, favoring configurations in which the spins lie in the $x-y$ plane, and also breaks the $Z_2$ invariance, favoring only one staggered chirality. The chirality is fixed by the assignment of the hopping phase of spin up electrons, shown in Fig.~\ref{fig:phi}(a). Further, for $\pi/4<\phi<3\pi/4$ the in-plane Heisenberg coupling changes sign, favoring ferromagnetic alignment of spins.  

These considerations lead us to investigate the Hartree-Fock energies of  the stripe, ferromagnetic, tetrahedral, and 120$^\circ$ ordered states. More specifically, we propose nine possible states: $120^{\circ}$ orders with two opposite staggered chiralities in the $x-y$ plane (120-xy-1, 120-xy-2) and in the $x-z$ plane (120-xz-1, 120-xz-2), a ferromagnetic state along the $z$ direction and in-plane (ferro-z, ferro-xy), a stripe state along the $z$ direction and in-plane (stripe-z, stripe-xy), and a tetrahedral state. These states are illustrated in Fig.~\ref{fig:phi}(b). 

In the Hartree-Fock treatment, the onsite interaction in Eq.~(\ref{eq:H}) is approximated as:
\begin{eqnarray}
    Un_{i\uparrow}n_{i\downarrow} \approx U &\sum_i & \langle n_{i\uparrow}\rangle n_{i\downarrow}+n_{i\uparrow}\langle n_{i\downarrow}\rangle -\langle n_{i\uparrow}\rangle \langle n_{i\downarrow}\rangle\nonumber\\
    & -&\langle S_i^+\rangle S_i^--\langle S_i^-\rangle S_i^++\langle S_i^+\rangle \langle S_i^-\rangle.
    \label{eq:HF}
\end{eqnarray}
Different magnetic states correspond to different position dependences of the averaged value $\langle S^z_{i}\rangle=\langle n_{i\uparrow}-n_{i\downarrow}\rangle$ and $\langle S_i^\pm\rangle=\langle c^\dagger_{i\uparrow}c_{i\downarrow}\rangle$/$\langle c^\dagger_{i\downarrow}c_{i\uparrow}\rangle$. For example, in the $x-y$ plane $120^{\circ}$ magnetic state, the averaged spin on each site $i$ follows $\langle S_i^x\rangle= m\cos(\vec{Q}\cdot\vec{R}_i+\theta),~\langle S_i^y\rangle= m\sin(\vec{Q}\cdot\vec{R}_i+\theta)$, where $m$ is the magnetization, $\theta$ is an arbitrary phase that determines where the spin points along the $x$ axis, and $\vec{Q}=(\pm4\pi/3,0)$ is the wave vector. Details of the Hartree-Fock Hamiltonian are given in Appendix \ref{app:meanfield}. 

We work in the canonical ensemble. In each magnetic state, the combination of spin order and $t^\sigma_{ij}$ determines a band structure. In the band structure, electron states are filled up to the chemical potential, and $m$ is determined by minimization of the total energy. In this way, the energy of each possible magnetic states is calculated, and the final ground state is chosen as the global minimum.

\section{Result: At half-filling}\label{sec:halffilling}
\begin{figure}[t]
  \includegraphics[width=0.9\columnwidth]{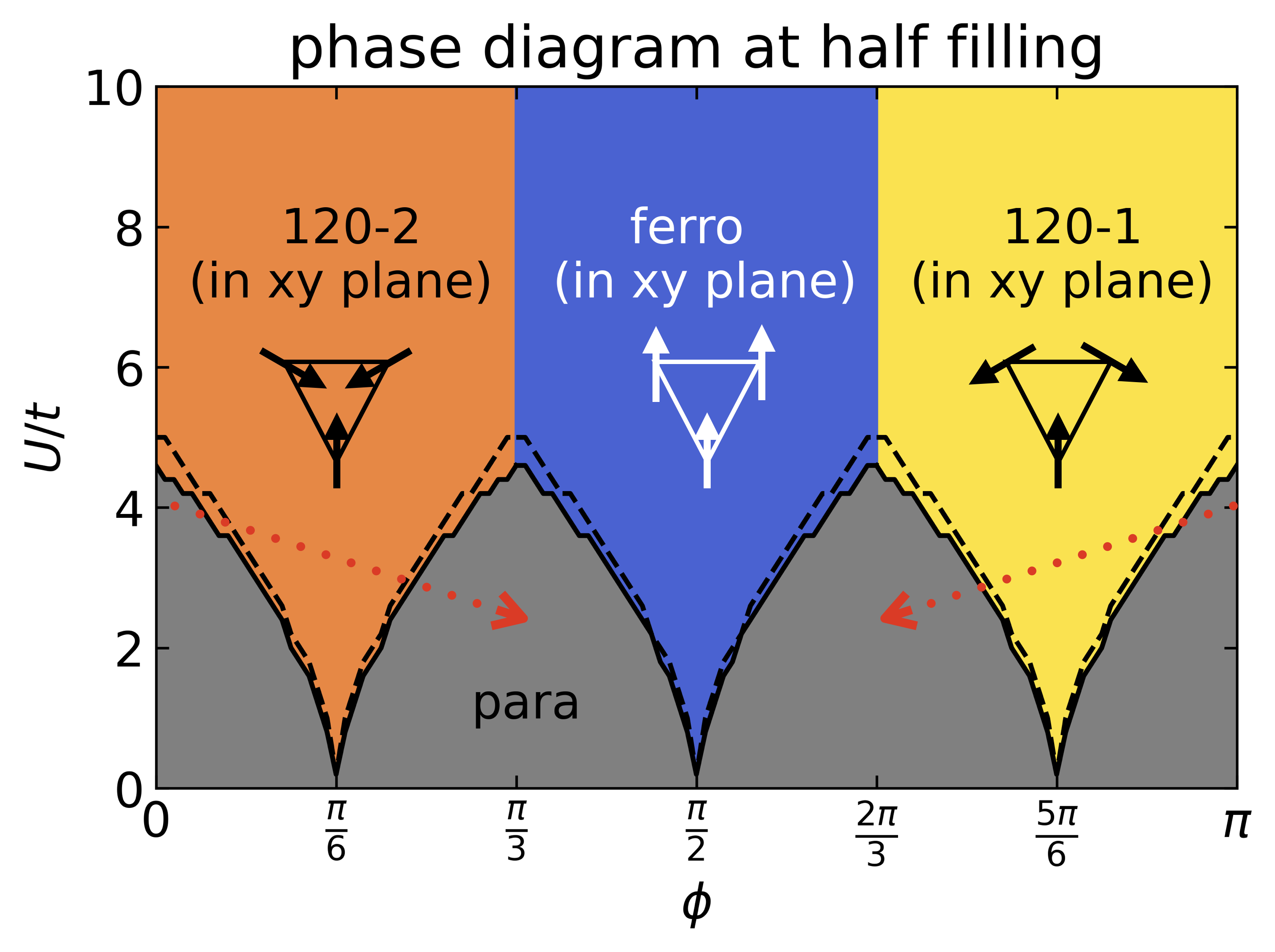}
  \caption{(a) Phase diagram at half filling. The solid black line marks the transition to the indicated magnetic order. The dashed black line marks the opening of the gap. The energy gap is calculated as the energy difference $\Delta E$ between the highest filled electron state and the lowest unfilled electron state.   The gap opening position is defined as the position where $\Delta E > 0.01$. The red arrow shows the parameter space trajectory followed when the displacement field is increased in experiments, which changes $\phi$ from $0$ to $\sim \pm \pi/3$ and increases $t$ thus decreasing $U/t$.}
  \label{fig:phase_half}
\end{figure}

Performing the Hartree-Fock calculation described in Section~\ref{sec:Method} at a carrier concentration $n=1$ per state, we find the phase diagram shown in Fig.~\ref{fig:phase_half}. The solid line marks the transition to magnetic order, and the dashed line marks the opening of a charge gap.  Three magnetic phases are found: the in-plane 120$^\circ$ phase (with chirality determined by the DM phase $\phi$), an in-plane ferromagnetic phase, and a paramagnetic phase. The magnetic phases are insulating over most of the $U$ values, but exhibit a small range of $U$ where metallic behavior and magnetic order coexist.  The sequence of magnetic phases occurring as $\phi$ is varied at large $U$ may be understood from the Heisenberg model shown in Eq.~(\ref{Heisenberg}) or more generally from the symmetries discussed above. The U-independence of the critical $\phi$ at which the magnetic order changes from  120$^\circ$ to ferromagnetic and the periodicity of the metal insulator phase boundary as $\phi\rightarrow \phi+\pi/3$ follow from the invariance of the spectrum under insertion of integer multiples of $\pi$ and the symmetries under $\phi \leftrightarrow -\phi$. We further note that the ferromagnetic state at $\phi=\pi/2$ is connected to the appropriate-chirality $120^\circ$ states by the space-dependent spin rotation discussed above. 

 Inclusion of further neighbor hopping terms in the band structure will break the symmetry. Second neighbor terms do not change the phase boundary but inclusion of third neighbor hopping will increase slightly the range of $\phi$ for which ferromagnetism is found and provide a weak $U$ dependence. Recent beyond Hartree-Fock studies of the model with $\phi=0$ suggest that while the 120$^\circ$ state found here is the large $U$ ground state, this state is separated from the paramagnetic metal state by an intermediate phase which has a charge  gap but lacks obvious long ranged magnetic order and is potentially a spin liquid \cite{wietek2021mott,PhysRevB.96.205130,PhysRevX.10.021042,chen2021quantum}.
 
 \begin{figure}[b]
  \includegraphics[width=0.9\columnwidth]{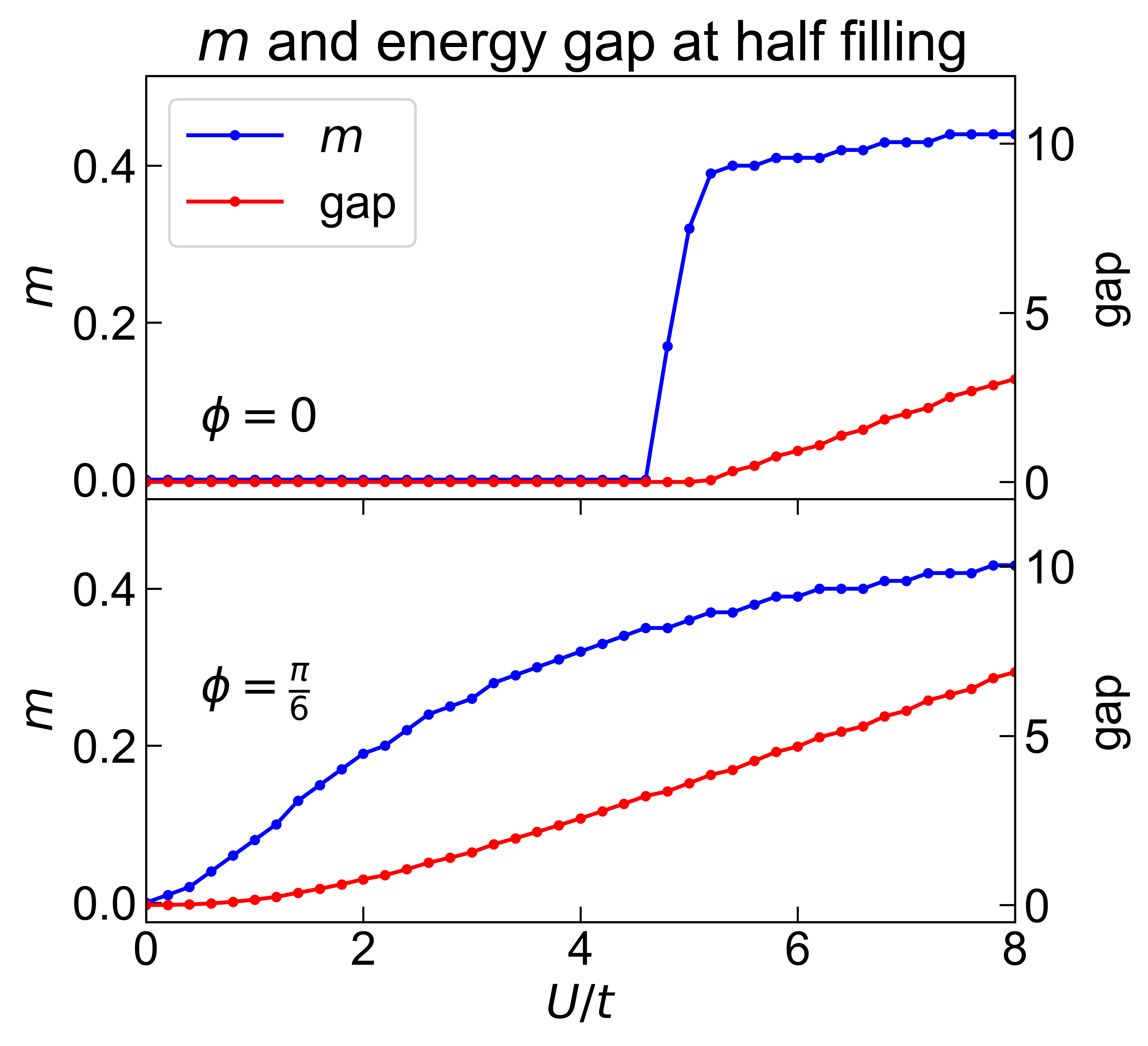}
  \caption{Magnetization $m$ and energy gap of 120$^\circ$ spiral order (120-xy-2) at $\phi=0$ and $\pi/6$ from Hartree-Fock calculation. The transition is found to be two-staged for almost all values of $\phi$ (a magnetic transition followed by a metal-insulator transition). At $\phi=\pi/6$, the magnetic transition coincides with the metal-insulator transition.}
  \label{fig:magandgap_half}
\end{figure}

Fig~\ref{fig:magandgap_half} shows the $U$ dependence of the magnetization $m$ and energy gap at two representative phases $\phi=0$ and $\pi/6$. The transition between paramagnetic metal and magnetic insulator exhibits a strong $\phi$ dependence, which is related to the van Hove singularity and nesting structure discussed in the next section. The transition is found to be two-staged for almost all values of $\phi$. As $U$ is increased a first transition to a magnetically ordered but still metallic state is observed, and then as $U$ is increased further a metal-insulator transition occurs. 
However it should be noted that the details of the narrow transition region between paramagnetic metal and antiferromagnetic insulator are complicated, with different incommensurately ordered magnetic metal states possibly occurring in a narrow $U$ range between the paramagnetic metal and antiferromagnetic insulator states \cite{krishnamurthy1990mott,Jayaprakash_1991}. More detailed investigation of these issues in the $\phi \neq 0$ model requires consideration of longer period incommensurate orders which is beyond the scope of this paper.  

Figure~\ref{fig:phase_half} shows that at $U\lesssim 5t$ the properties are reentrant as $\phi$ is varied, with a metallic phase at $\phi=0$ giving way to an insulating phase for $\phi$ near $\pi/6$ then evolving back to a metallic phase as $\phi$ is increased beyond $\pi/6$. Experimentally, $\phi$ is increased from 0 by varying the ``displacement field'' $D$ (interlayer potential difference), which also increases $|t|$, so the experimental system explores a trajectory shown qualitatively by the red dashed lines, going from metallic at $D$=0 through insulating and back to metallic as $D$ is increased, indicating that the interaction in \twsetwo~is at an intermediate level as has previously been noted \cite{Wang:2020us,ghiotto2021quantum}.

\begin{figure}[htbp]
  \includegraphics[width=0.9\columnwidth]{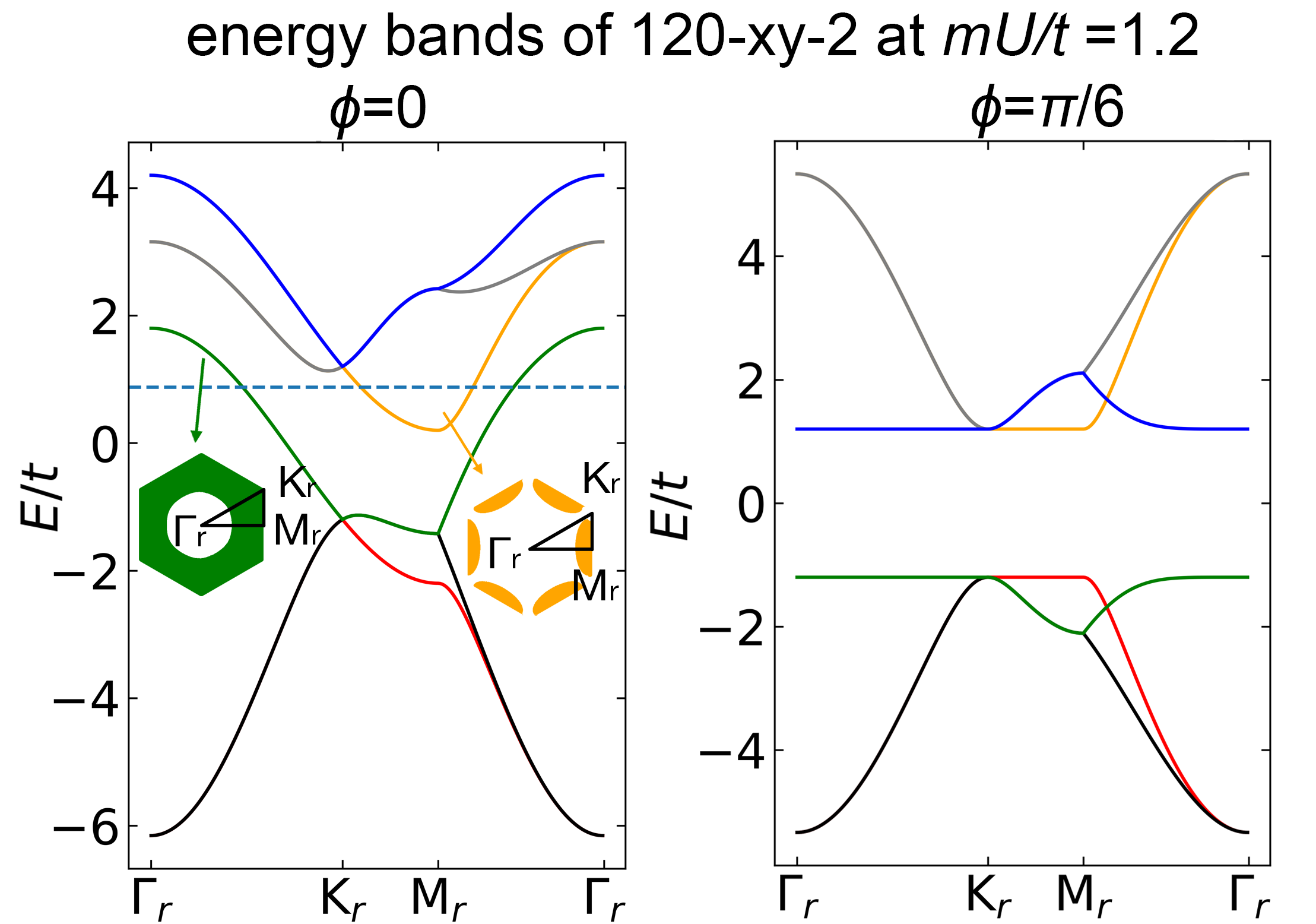}
  \caption{Energy bands of 120-xy-2 at $\phi$=0 and $\pi/6$ at an intermediate interaction U. The band is plotted along lines in the folded zone of the $120^{\circ}$ three sublattice spiral order, which is one third of the original moir\'e Brillouin zone. The green zone shows the hole pocket of the lower band and the yellow zone shows the electron pockets of the upper band at $\phi$=0. In the left panel, the dashed blue line indicates the chemical potential at half filling. In the right panel, any chemical potential in the band gap corresponds to half filling.}
  \label{fig:pocket}
\end{figure}
 
To clarify the nature of the metal-insulator transition in this model, we plot in Fig.~\ref{fig:pocket} the band structure in the magnetic Brillouin zone for $\phi=0$ and $\phi=\pi/6$ at a moderate $U$. At $\phi=0$ we see that the band structure consists of a hole pocket centered at $\vec{\Gamma}$ and electron pockets centered at the $\vec M$ point. As the interaction is increased the energy separation between the lower and upper bands increases, decreasing the sizes of the electron and the hole pockets. As $\phi$ is varied at fixed U, the bands flatten and separate, similarly leading to a metal-insulator transition. At $\phi=\pi/6$, the perfect nesting, which will be discussed in next section, leads to flat bands in the magnetic zone and a metal-insulator transition coincident with the magnetic transition, which indeed occurs at $U=0$.

\begin{figure*}[ht]
  \centering 
  \includegraphics[scale=0.3]{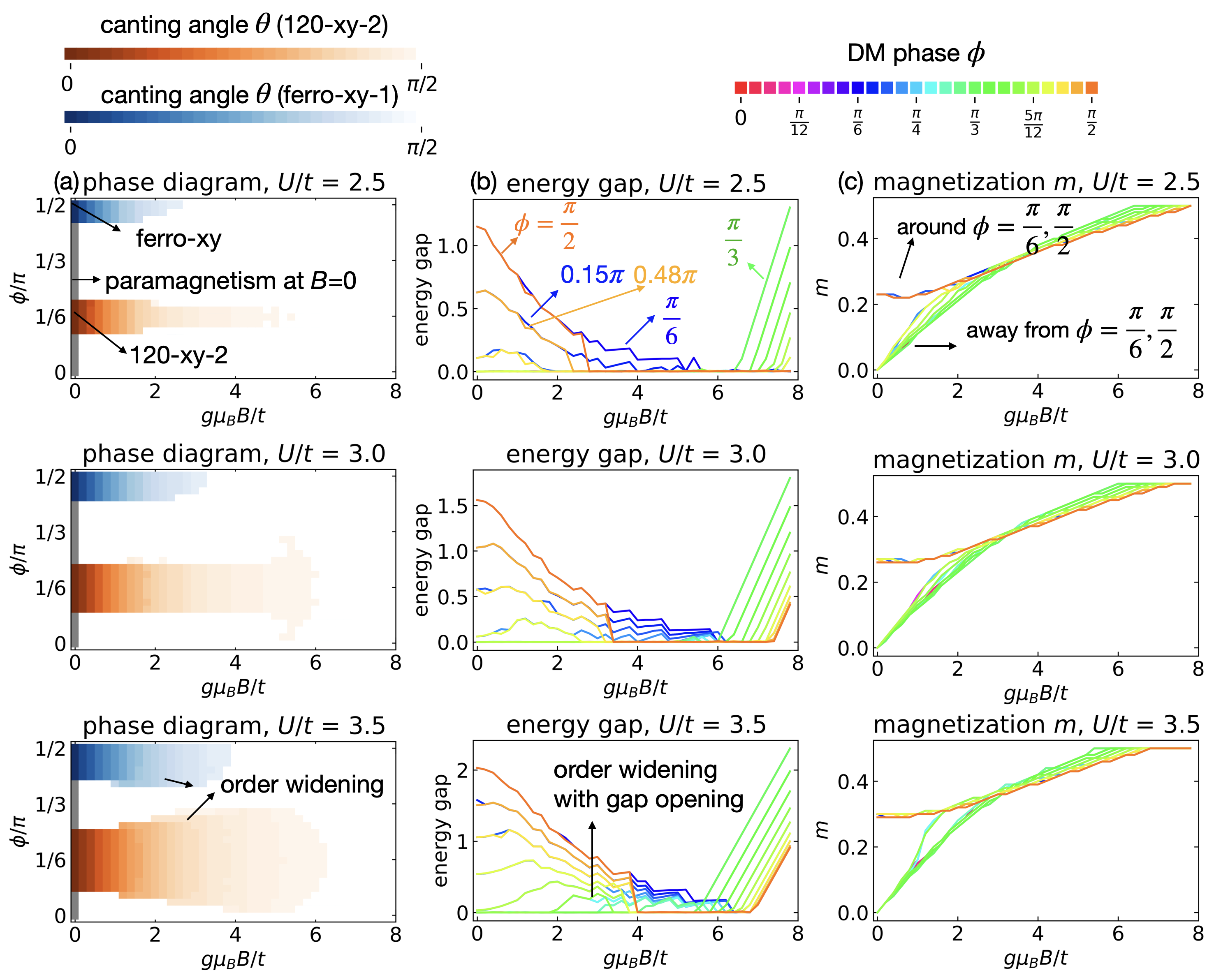}
  \caption{(a) Phase diagram, (b) energy gap and (c) magnetization at half filling for $0\leq\phi\leq\pi/2$ for several U values. (a) In the phase diagram, gray represents the paramagnetism. The intensity of the blue and orange color indicates the canting angle $\theta$ of the magnetic order. Dark orange and dark blue represent the 120-xy-2 order and ferro-xy correspondingly, and the white regions have no xy moment. (b,c) In the energy gap and magnetic order plots, the colors represent the corresponding DM phase $\phi$. For $g\approx 10$ in tWSe2 with bandwidth around 100meV$\approx 10t$, $g\mu_BB/t=1$ corresponds to $B\approx 17$ T.}
  \label{field}
\end{figure*}

We next discuss the effect of a Zeeman magnetic field. A Zeeman field perpendicular to the plane generally will cause the in-plane magnetic order $m(\cos\vec{Q}\cdot \vec{R}_i, \sin\vec{Q}\cdot \vec{R}_i,0)$ to gradually cant towards the $z$ direction with a canting angle $\theta$. Again we use the Hartree-Fock treatment. We assume the in-plane magnetic state becomes $m(\cos\theta\cos\vec{Q}\cdot \vec{R}_i,\cos\theta\sin\vec{Q}\cdot \vec{R}_i,\sin\theta$) and then calculate the ground state by minimizing the energy with respect to $m$ and $\theta$. Results are summarized in Fig.~\ref{field} at several intermediate $U$. Since the $g$ factor in \twsetwo~has a relatively large uncertainty, and Hartree-Fock is good at qualitatively capturing the changes, the picture is plotted for a wide range of the magnetic field.

At small $U$ and $B=0$, magnetic orders are found around $\phi =\frac{\pi}{6}$ (120-xy-2) and $\frac{\pi}{2}$ (ferro-xy) with non-zero energy gaps. Other regions are paramagnetic. As the magnetic field increases, the original in-plane magnetic order first increases the canting angle $\theta$ without changing $m$ much, then increases both $\theta$ and $m$ until they reach a maximum, and the energy gap gradually decreases to zero. In paramagnetic regions, after the field is turned on, spins will quickly align to the $z$ direction with zero energy gap. If the magnetic field is extremely large ($g\mu_BB/t>6$), there will be a new gap opening, due to the splitting of the spin up and spin down bands.

For intermediate $U$, applying a $z$ direction magnetic field can produce xy order at $\phi$ values where there is no order at $B=0$. For example, at $U/t=3.5$, the left panel (``phase diagram'') clearly shows that the $\phi$ range of magnetic orders widens as $B$ is increased. For this interaction strength, increasing $B$ can produce an energy gap in a finite range of $B$ (see curve marked in Fig.~\ref{field}). Thus, by tuning the displacement field and the magnetic field, there could be some gap opening and closing, related to the appearance of canted magnetic order.

To conclude this section we consider briefly some extensions of our results. The Hartree-Fock theory we present here is restricted to classically definable magnetic orders. The putative spin liquid phases indicated by numerics for the model with $\phi=0$ are not captured by our formalism. Understanding how such phases evolve as $\phi$ is varied is an important open problem. Our restriction to only first neighbor hopping fixes the critical value of $\phi$ where the van Hove point coincides with the half filled Fermi surface to be $\phi=(2n+1)\pi/6$. Inclusion of further neighbor hopping would shift the critical $\phi$ and would remove the perfect nesting, but the van Hove singularity remains, and the qualitative behavior is unchanged.

\section{General fillings }\label{sec:phase diagram all}

\begin{figure}[htb]
  \includegraphics[width=0.8\columnwidth]{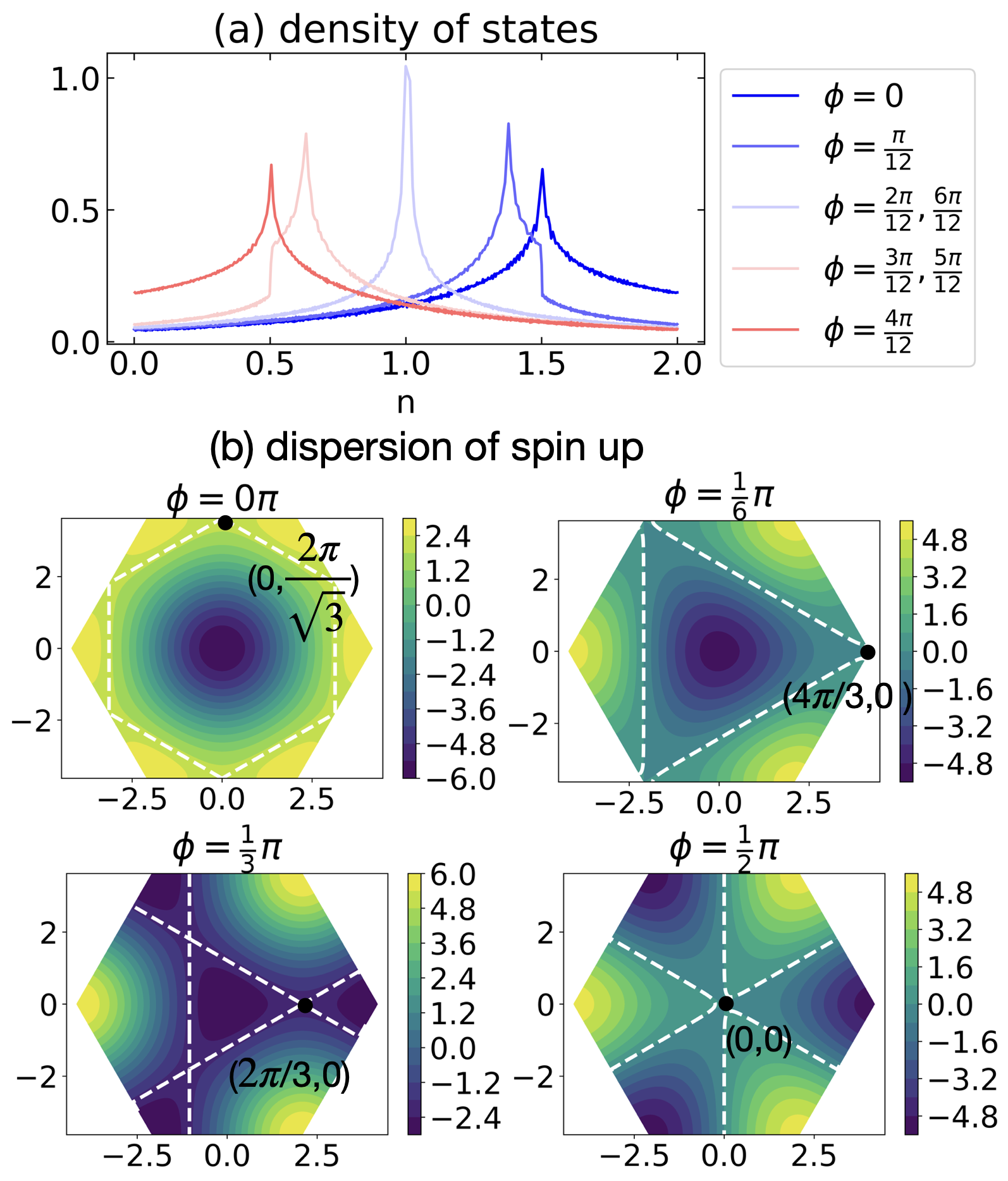}
  \caption{(a) Density of states versus filling calculated at $U=0$. Colors represent different choices of $\phi$. (b) Dispersion of spin up electrons in the moir\'e Brillouin zone calculated at $U=0$. The dashed white line shows the constant energy surface that intersects the energy saddle point. The labelled point in each plot indicates one of the van Hove locations. }
  \label{fig:DOS}
\end{figure}


When the density $n\neq 1$, charge fluctuations mean that Hartree-Fock calculations becomes less reliable. A complicated variety of commensurate and incommensurate ordered phases along with regions of phase separation have been reported for the model without spin orbit coupling \cite{HF_all_filling,pasrija2016noncollinear}, but the effect of beyond-Hartree-Fock fluctuations has not been established. In this section, we present a qualitative discussion focussed  on the $\phi$ dependent weak coupling instabilities, which are controlled by nesting and van Hove singularities, for which a Hartree-Fock based approach is more reliable.

For general $\phi$  the spin up and spin down Fermi surfaces do not coincide. The van Hove (saddle point) singularity, which is generically present in two dimensional band structures, lies at a band filling which varies smoothly with displacement field $\phi$, and is visible as a divergence in the density of states plots, as shown in Fig.~\ref{fig:DOS}(a). We extract the numerically calculated density $n_{vHs}$ where the Fermi surface intersects with the van Hove singularity, and find that the numerically calculated $n_{vHs}$ is well fitted by $n_{vHs}\approx\cos(3\phi)/2+1$.

Fig.~\ref{fig:DOS}(b) shows the electron energy dispersion for spin up at zero interaction for different values of DM phase $\phi$, along with the energy isosurface that passes through the van Hove points. For the nearest neighbor model studied here, the energy contour passing through the van Hove points has flat regions, leading to nesting. The combination of density of states divergence and nesting destabilizes the paramagnetic metal  state at infinitesimal $U$ if the density $n=n_{vH}(\phi)$ is chosen so that the Fermi energy passes through the van Hove point.

\begin{figure}[htbp]
  \includegraphics[width=0.9\columnwidth]{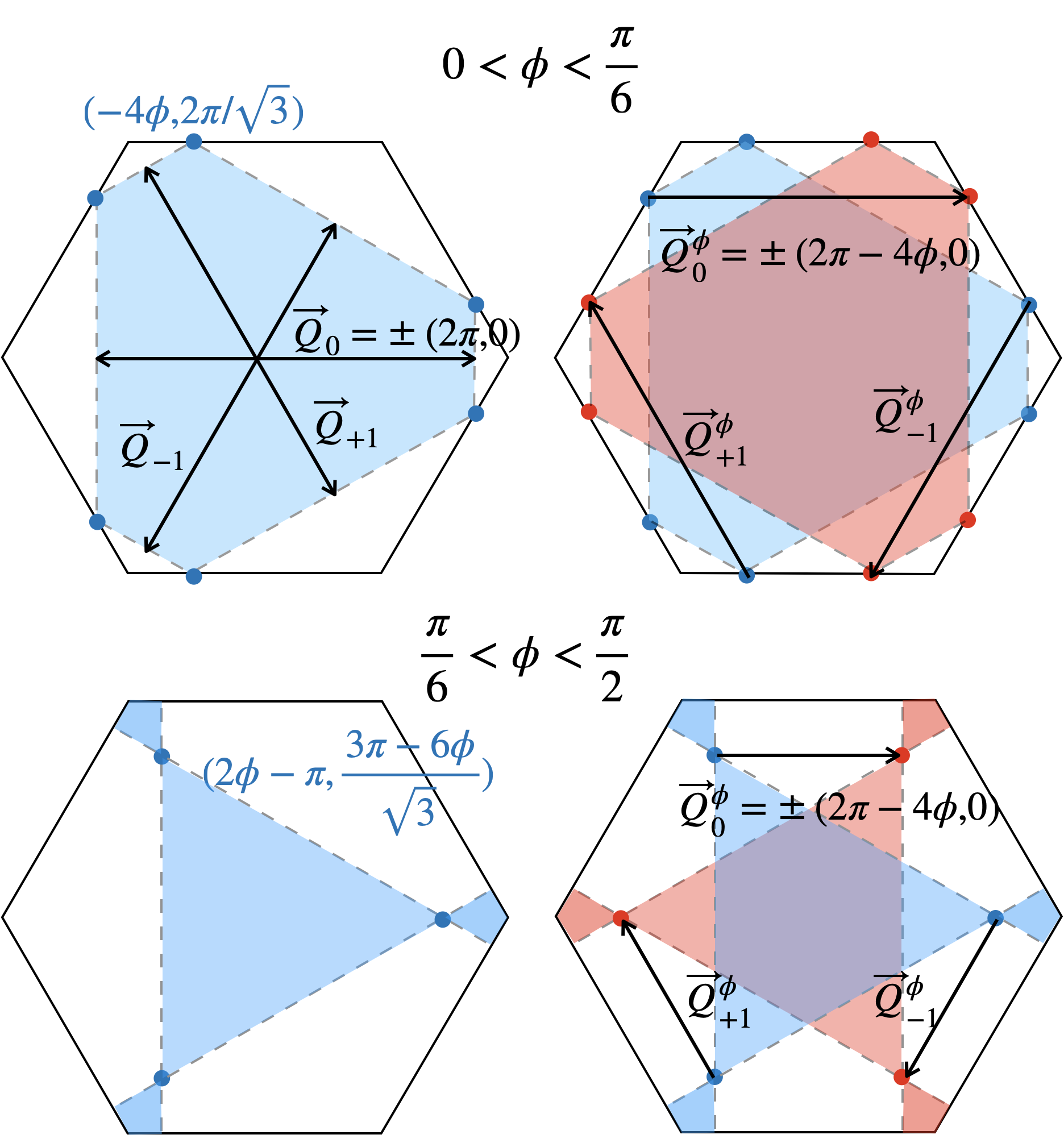}
  \caption{An illustrative sketch of occupied states. Blue (Red) area represents the occupied Brillouin zone of spin up (down). Blue (Red) dots are the van Hove locations of spin up (down). The arrow indicates the wavevector that connects the perfect nesting electrons between the same spins ($\vec{Q}_{0,\pm 1}$) and between spin up and down ($\vec{Q}^\phi_{0,\pm 1}$).}
  \label{fig:vHs_location}
\end{figure}

In the nearest neighbor hopping model considered here, the van Hove points at $\phi=0,\pi/6,\pi/2$ are special. At $\phi=0$ the van Hove points of the spin up and spin down Fermi surfaces coincide and the van Hove points are at the $\vec{M}$ and $\vec{M}^\prime$ points of the Brillouin zone. As $\phi$ is increased from $0$ the van Hove points shift asymmetrically away from the $\vec{M}/\vec{M}^\prime$ points while remaining at the zone boundary. At $\phi=\pi/6$ the van Hove points coalesce into a third order singularity at the $\vec{K}$ (spin up) or $\vec{K}^\prime$ (spin down) points; for $\pi/6<\phi<\pi/2$ the van Hove points move inwards along the $\vec{\Gamma}-\vec{K}/\vec{K}^\prime$ lines, coalescing again at a third order singularity at the $\vec{\Gamma}$ point at $\phi=\pi/2$. The particular numerical values of $\phi=0,\pi/6,\pi/2$ at which the three special conditions occur are particular to the nearest neighbor only model; use of a more general dispersion will change the  values of displacement field  at which the three special van Hove points occur and the band fillings  at which they lie at the Fermi surface, but the special van Hove points will in general exist.

The weak coupling physics can be understood via consideration of the saddle point action
\begin{equation}
    S[\{\vec{m}\}]=\text{Trln} \left[G_{0,\sigma}^{-1}+\vec{m}_i\cdot\vec{\sigma}\right]+\sum_i\frac{\vec{m}_i^2}{U}
    \label{action}
\end{equation}
where $G_0^{-1}=\partial_\tau-t_{ij}^\sigma-\mu$ is the noninteracting Green function, $\vec{\sigma}$ is the Pauli matrix, and $\vec{m}_i$ is a Hubbard-Stratonovich field proportional to the expectation value of the spin on site $i$. We study the free energy of static configurations of the $\vec{m}_i$, evaluating the trace term by expansion in $\vec{m}_i$. After the Fourier transformation, the second order term is
\begin{equation}
    F^{(2)}_\phi=\sum_{\vec{Q}}\frac{1}{2}\sum_{a,b}m^a_{\vec{Q}}\left(U^{-1}\delta_{ab}-\chi^{ab}_0(\vec{Q})\right)m^b_{-\vec{Q}}
    \label{eq:S2}
\end{equation}
where $m^a$ is the component of $\vec{m}$ in Cartesian direction $a$, and we introduce the susceptibility coefficient $\chi^{ab}_0$ 
\begin{equation}
    \chi^{ab}_0(\vec Q)=\text{Tr}\left[\bm{\sigma}^a\mathbf{G}_0(\vec k)\bm{\sigma}^b\mathbf{G}_0(\vec k+ \vec Q)\right]
\end{equation}
where $\mathbf{G}_0$ is the (diagonal) Green function matrix in spin and momentum space, and the trace is over spin and momentum indices.

The action is dominated by the susceptibility with perfect nesting wavevectors. From Fig.~\ref{fig:vHs_location}, we see that in the nearest neighbor model we consider here, there are two kind of nesting vectors: $\vec{Q}_{0,\pm 1}$ that connects the the Fermi surface of same spins, and $\vec{Q}_{0,\pm 1}^{\phi}$ that connects the the Fermi surface from spin up to spin down. $\vec{Q}_0^{\phi}=(2\pi-4\phi,0)$, and $\vec{Q}_{\pm 1}^{\phi}=R_{\pm2\pi/3}\vec{Q}_0^{\phi}$ is obtained from $\vec{Q}_0^{\phi}$ by rotations of $\pm$ $2\pi/3$ about $z$ axis.  For $\phi\neq 0,\pi/6,\pi/2$, the wave vector $\vec{Q}_{0,\pm1}^{\phi}$ both connects the nesting surfaces and the van Hove points. The result is $log^2$ divergences in susceptibilities, corresponding to shifting an electron by $\vec{Q}$ and flipping spin up to spin down. We also see that the wavevectors $-\vec{Q}_{0,\pm1}^{\phi}$ do not connect van Hove points or flat regions of Fermi surface from spin up to down. Further for generic $\phi\neq0,\pi/6,\pi/2$, $\vec{Q}_{0,\pm 1}$ does not connect the van Hove points, and the flat regions of Fermi surface connected by $\vec{Q}_0$ are of different lengths, leaving only a $log$ divergence with a smaller coefficient. The result is that the dominant terms in $\chi^{ab}$ are susceptibilities  $\chi^{+-}_0(\vec{Q}^\phi_{0,\pm1})=\chi_0^{-+}(-\vec{Q}^\phi_{0,\pm1})$, 
implying linear instabilities to the three stripe spiral orders  with  spin pattern
\begin{equation}
    S^x_l(\vec{R})=S_l\cos\left(\vec{Q}_l\cdot\vec{R}+\theta_l\right);~S^y_l(\vec{R})=S_l\sin\left(\vec{Q}_l\cdot\vec{R}+\theta_l\right);
\end{equation}
where  $\vec{Q}_l=\vec{Q}^\phi_{0,\pm1}$, and $\theta_l$ determines the locations where the spin points along $x$ in the spiral with wavevector $\vec{Q}_l$. The corresponding order parameters are most conveniently written as $\left<m_l^+\right>=\left<m_l^x+im_l^y\right>=S_le^{i(\vec{Q}_l\cdot\vec{R}+\theta_l)}$.

At the quadratic level the three spiral directions are equivalent. At quartic level, expansion of the action gives terms 
\begin{eqnarray}
    F^{(4)}_\phi&=&\frac{\beta_1} {4T^2}\sum_l\left(S^+_{Q_l}S^-_{-Q_l}\right)^2+\nonumber \\
    &&\frac{\beta_2 }{4T}\sum_{l\neq s}\left(S^+_{Q_l}S^-_{-Q_l}\right)\left(S^+_{Q_s}S^-_{-Q_s}\right)
    \label{eq:S4}
\end{eqnarray}
where $\beta_{1,2}$ are constants, and the factors of $T$ arise because if all four $S$ share the same wavevector then the corresponding diagram is $\sim G(p)^2G(p+Q)^2$ and is strongly divergent at the nesting wavevector while if two different wavevectors are involved then at most one pair of $G$ can be nested.

Minimizing Eqs.~(\ref{eq:S2}), (\ref{eq:S4}) we find that the free energy minimum corresponds to three $x-y$ plane spirals, along the three wavevectors $\vec{Q}_{0,\pm1}^\phi$, each of equal amplitude, and with phases $\theta_l$ that are arbitrary. In the nearest neighbor hopping model considered here the trio of spiral states fully gaps the Fermi surface, leading to an insulator at the corresponding nesting density. As the coupling strength is increased, commensurability energies come in to play and we expect that the physical state corresponds to regions of commensurate order with discommensurations that can trap charge carriers, in analogy to the stripe states found in the square lattice Hubbard model \cite{PhysRevB.40.7391,PhysRevB.44.943,machida1989magnetism,kato1990soliton}. Therefore, there will be an insulator-metal transition as interaction increases. If further neighbor hopping is included, the perfect nesting is spoiled and regions of the Fermi surface could remain ungapped at weak coupling. 

We now consider the three special cases, beginning with $\phi=0$ at $n=1.5$. For this $\phi$ the spin up and spin down Fermi surfaces coincide.  The van Hove points are at the M and M' points of the moire Brillouin zones (density $n=1.5$), and the nesting vectors are at $\vec{Q}_0=\pm(2\pi,0)$ and $\vec{Q}_{\pm1}=\pm2\pi(-\frac{1}{2},\pm\frac{\sqrt{3}}{2})$. The coincidence of spin up and spin down Fermi surfaces mean that the nearest neighbor model has $SU(2)$ spin invariance, seen here in the fact that the spin up Fermi surface nests with both the spin up and spin down Fermi surfaces, and $\pm \vec{Q}_{l}$ are both nesting vectors. The wavevector $\vec{Q}_0=(2\pi,0)$ means that the spiral has vanishing pitch, so the state is a collinear stripe of form shown in Fig.~\ref{fig:phi}(b). An analysis similar to that sketched in Eqs.~(\ref{eq:S2}), (\ref{eq:S4}) gives an $SU(2)$-invariant theory with quadratic term $\sum_l\vec{S}(\vec{Q}_l)\cdot\vec{S}(\vec{Q}_l)$ and dominant quartic term $\sum_l \left(\vec{S}(\vec{Q}_l)\cdot\vec{S}(-\vec{Q}_l)\right)^2$ so that at this level  the free energy is minimized by three equal amplitude collinear stripes, with  orthogonal spin directions. As noted in Ref.~\cite{martin2008itinerant}, sixth order terms in the free energy then fix the phase between the three stripes, inducing  a chirality. The chiral state is disfavored by the spin orbit coupling appearing if further neighbor interactions are considered. 

As $\phi$ increases from 0, we see from Fig.~\ref{fig:vHs_location} that for the same-spin nesting, the length of one nesting edge decreases continuously to zero, and the nesting vector is separated from van Hove locations, implying a rapid decrease in the strength of the divergence. On the other hand, the nesting vector of opposite spins still connects the van Hove locations, implying the rapid development of an easy plane anisotropy.  

At $\phi=\frac{\pi}{6}$, the same-spin nesting vectors disappear, and the spin up (down) van Hove points merge at the high symmetry points $\vec K$ ($\vec K'$), producing a cubic van Hove singularity ($\epsilon_k\sim k_x^3-3k_xk_y^2$). Such high order van Hove singularity will lead to a power-law-divergent density of states and $\chi^{\pm}$, implying a stronger tendency towards order \cite{shtyk2017electrons,yuan2019magic,bi2021excitonic,isobe2019supermetal}. The ordering wavevector $\vec{Q}^\phi_{0}=(\frac{4\pi}{3}, 0)$  is equivalent to its $C_3$ rotations (up to a reciprocal lattice vector), so the three spiral states merge into one $120^\circ$ spiral in-plane order with a definite staggered chirality. This state will gain substantial commensurability energy, and it is likely that the general $\vec{Q}$ states found at other values of the displacement field will evolve into defected versions of the $120^\circ$ state as the interaction is increased.

For $\frac{\pi}{6}<\phi<\frac{\pi}{2}$, the van Hove points move to the interior of the zone along the $\vec{\Gamma}$-$\vec K$ ($\vec K'$) line, and the opposite-spin nesting continues to exist at wavevectors $\vec{Q}^\phi_{0,\pm\frac{2\pi}{3}}$ (see Fig.~\ref{fig:vHs_location}). At $\phi=\frac{\pi}{2}$, all van Hove singularities merge into the third order singularity at $\vec{\Gamma}$, there is no nesting, and the predicted magnetic state is ferromagnetic.

\begin{figure}[ht]
  \includegraphics[width=0.9\columnwidth]{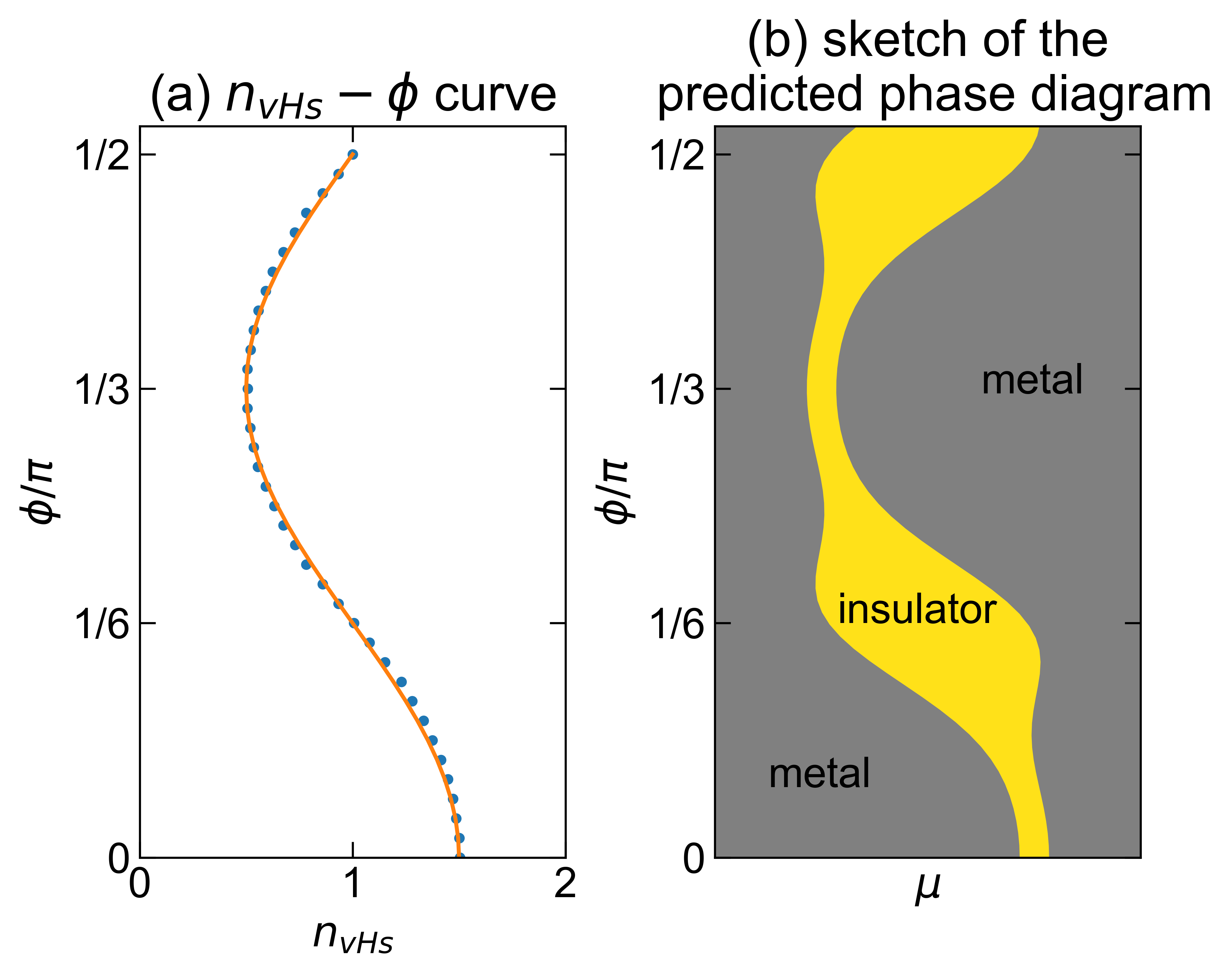}
  \caption{(a) $n_{vHs}-\phi$ curve for $0<\phi<\frac{\pi}{2}$ at $U$=0. The blue points are extracted from the numerical density of states calculation. $n_{vHs}$ is the density filling where the Fermi surface intersects with the van Hove singularity. The orange line is an empirical formula $n_{vHs}\approx\cos(3\phi)/2+1$ that fits the numerical calculation well. (b) Sketch of the predicted phase diagram with only nearest neighbor hopping in the weak coupling limit. }
  \label{fig:predicted_phase}
\end{figure}

To summarize, for weak coupling, the nearest neighbor hopping model predicts magnetically ordered insulating states along the line in the density-$\phi$ plane shown in Fig.~\ref{fig:predicted_phase}(a). For most values of $\phi$ the insulating states correspond to a triple of $x-y$ spirals with a fixed staggered chirality ($\phi$ dependent wavevector), but at $\phi=0$ the state is the chiral tetrahedral ordered state and at $\phi=\pi/2$ the state is an $x-y$ ferromagnet. If further neighbor hopping is included, then the incomplete nesting means the very weak coupling state is a magnetic metal. At general $\phi$, the incommensurate value of the spiral wavevector and the absence of any energetic term fixing the relative phases of the spirals means that the state is very susceptible to fluctuations. Also, as $U$ is increased other states may occur. For example, at $n$ near 1.5 and $
\phi=0$, Hartree-Fock calculation indicates that the tetrahedral state is replaced by a ferromagnetic state as $U$ is increased above a critical value $\sim 3.5|t|$ \cite{pasrija2016noncollinear}. For $\phi$ closer to $\pi/6$ the commensurability energy gain of the simple $\vec{Q}=(4\pi/3,0)$ 120$^\circ$ spiral state suggests that at intermediate and large U the state is likely to be a defected 120$^\circ$ state. However, if weak coupling versions of the material can be implemented, the lines of phase transition noted here should be observed. In Fig.~\ref{fig:predicted_phase}(b), we show a sketch of the predicted phase diagram for the nearest neighbor hopping model, where the insulator behavior could be found for general $\phi$, with the wavevector of the insulating spiral state varying.




\section{Conclusion}\label{sec:summary}
\begin{figure}[ht]
  \includegraphics[width=0.9\columnwidth]{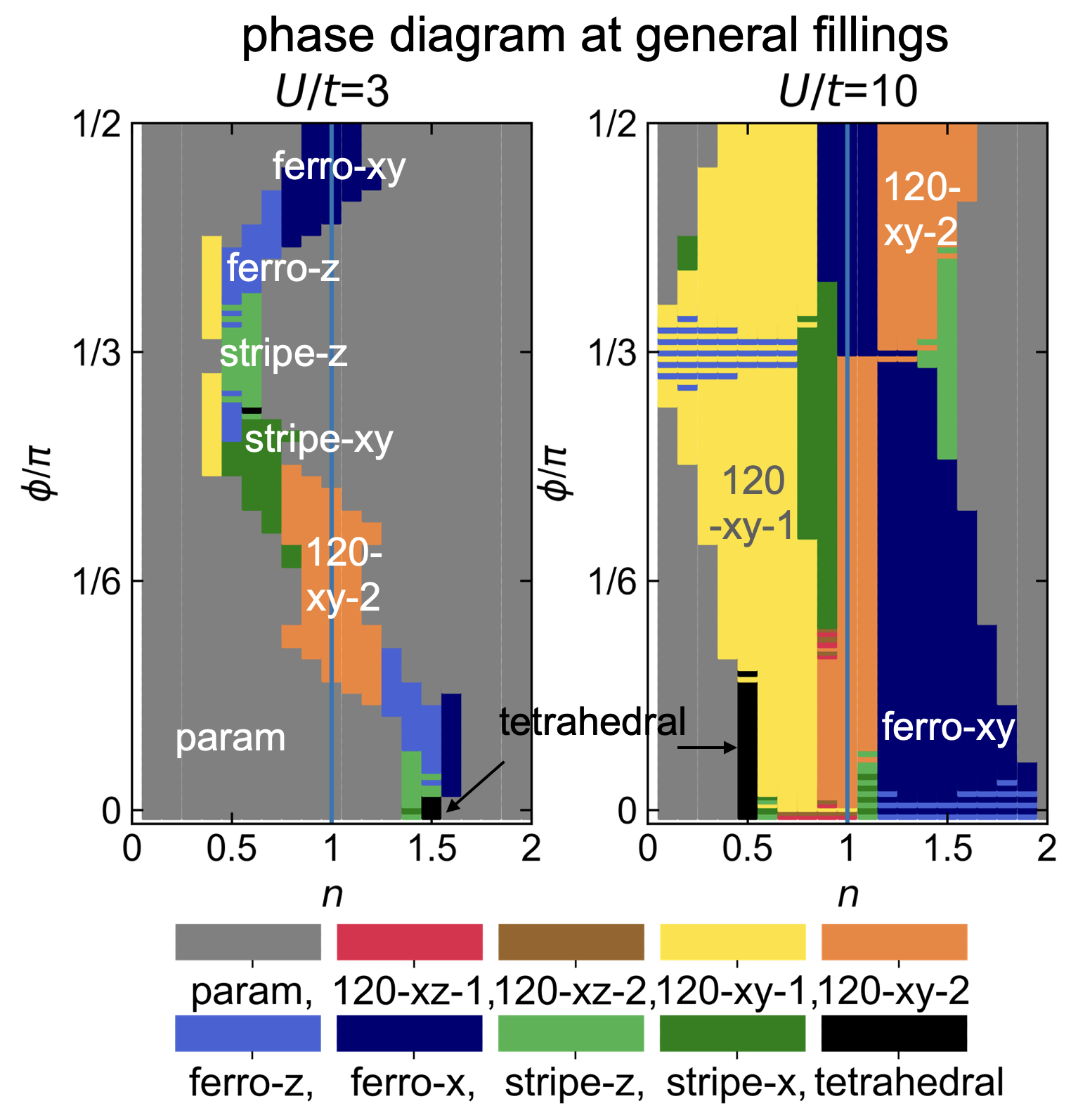}
  \caption{Hartree-Fock phase diagram at general fillings at weak and strong couplings with nine commensurate orders considered. Each color represents a different magnetic order.``xy'' indicates that the magnetic order is in the $x-y$ plane, and ``z'' represents the $z$ direction. Regions filled by more than one color are viewed as degenerate regions, where the energy difference between the two magnetic orders is smaller than $10^{-3}$ from numerical calculations. }
  \label{fig:phase_U}
\end{figure}

In this work, we present a comprehensive Hartree-Fock study of the moir\'e Hubbard model believed to represent the low energy physics of twisted WSe$_2$ and related materials. The new feature of the moir\'e Hubbard model is strong tunable spin orbit coupling, leading to a magnetic easy-plane anisotropy and highly tunable  van Hove singularity. The $g$-factor parameterizing the electron-spin interaction is large and anisotropic. The $O(2)$ rather than $SU(2)$ spin symmetry of the generic model is expected to reduce the importance of quantum fluctuations, increasing the parameter ranges where the orders found in the Hartree-Fock calculation are stabilized, and also ensuring that magnetic phases found at $T=0$ will persist for a range of nonzero temperatures.

At half filling, we find that for $U$ greater than a critical value $\sim 5|t|$, the model is magnetically ordered with a charge gap at all $\phi$. The predicted magnetic order depends on $\phi$, with regions of 120$^\circ$ spiral and regions of ferromagnetism. The ferromagnetic regions occur at $\phi$ values corresponding to displacement fields at the edge of what can be realized experimentally.  At smaller $U$, a reentrant phase diagram is found, with a metallic phase at $\phi=0$ giving way to an insulating phase for $\phi$ near $\pi/6$ and then reverting to a metallic phase. Experimental results for devices with twist angle $\sim4-5^\circ$ indicate a similar reentrance, placing these devices in the intermediate coupling regime.  Smaller twist angles would enlarge the unit cell \cite{PhysRevResearch.2.033087}, decreasing both the hopping and the interaction terms. Since the hopping decreases faster, the net effect of a smaller twist angle is to increase $U/t$, pushing the system into the strong coupling regime. 

At general band fillings and interaction strengths, previous Hartree-Fock studies of the $SU(2)$ invariant model find a intricate phase diagram, with regions of stripes, phase separation, and defected commensurate phases, all occurring at general interaction strengths and carrier concentrations. Fig.~\ref{fig:phase_U} shows our Hartree-Fock phase diagram as a function of displacement field, where only nine commensurate orders are considered. In the weak coupling limit, if incommensurate orders are included in the nearest neighbor model, the insulating behavior should be found along the van Hove density and DM phase $n_{vHs}-\phi$ curve, due to the van Hove singularities and perfect nesting, as shown in Fig.~\ref{fig:predicted_phase}. When the density is away from half filling, as interaction increases, it is likely that a commensurate-incommensurate transition will occur in the magnetism, so that away from half filling  the incommensurate insulating phases would be replaced by commensurate magnetic metal phases.

A particularly interesting feature of the phase diagram of the SU(2) invariant triangular lattice Hubbard model is that at half filling the large $U$ 120$^\circ$ phase is separated from the low U fermi liquid metallic phase by an intermediate phase occurring for $U/t\sim9$ that has no obvious long ranged order and has been interpreted as a spin liquid \cite{PhysRevB.96.205130}, though the identification is not yet confirmed. The evolution of this potential  spin liquid state as $\phi$ and carrier concentration are varied is an interesting open problem.

In conclusion we further observe that the Moire Hubbard model studied here, this model is an approximate description of  emergent low energy properties of a richer and more complex system. For example, the microscopic origin of spin up and down states in terms of the two valleys of the top and bottom layer, along with the strong spin orbit coupling,  raises the possibility of anomalous  electron-phonon interactions. The study of these and related phenomena are important open questions for future research. 

\begin{acknowledgments}
We thank Augusto Ghiotto, Abhay Pasupathy, and Cory Dean for discussions on experimental results. We thank Antoine Georges, E. Miles Stoudenmire, Martin Claassen and, especially, Alex Wietek for fruitful discussions. J.C., J.Z. and A.J.M acknowledge support from the NSF MRSEC program through the Center for Precision-Assembled Quantum Materials (PAQM) - DMR-2011738. The Flatiron Institute is a division of the Simons Foundation.
\end{acknowledgments}

\appendix
\section{Band Structure}\label{app:band_structure}
The band structure was originally calculated using density functional theory (DFT) in Ref.~\cite{Wang:2020us}. It can be understood and qualitatively modelled based on a low-energy continuum model \cite{wu2019topological}.

In the monolayer WSe$_2$, the two valleys $\vec K_0$ and $\vec K'_0$ are dominated by opposite spins and are related by time reversal symmetry, as discussed Section \ref{sec:Model}. Here we focus on the $\vec K_0$ valley, which is dominated by spin up. Using $\vec{k}\cdot \vec{p}$ theory \cite{Korm_nyos_2015}, the single layer Hamiltonian of the top valence band can be approximated as
\begin{equation}
    h({\vec{k}-\vec{K}_0})=-(\vec{k}-\vec{K}_0)^{2} / 2 m^\ast + \tau\mathcal{C}|\vec{k}-\vec{K}_0|^{3} \cos (3 \alpha_{\vec{k}}),
    \label{eq:Hmonolayer}
\end{equation}
where $m^\ast$ is the effective mass, $\tau=1$ indicates the valley $K_0$ ($-1$ for $K'_0$), and $\alpha_{\vec{k}}=\arctan\frac{(\vec{k}-\vec{K}_0)|_y}{(\vec{k}-\vec{K}_0)|_x}$. The term $\mathcal{C}|\vec{k}-\vec{K}_0|^{3} \cos (2 \alpha_{\vec{k}})$ preserves the $C_3$ symmetry of the $\vec{K}_0$ point of the monolayer and was not explicitly written in Ref. \cite{wu2019topological}. In the bilayer, this term preserves the $C_3$ rotation symmetry and $C_{2x}$ symmetry and protects a band degeneracy along certain high symmetry lines in the Brillouin zone. When $\mathcal{C} = 0$, the monolayer dispersions entering the bilayer model have an $O(2)$ rotation invariance which becomes an emergent inversion symmetry ($E_\sigma(\vec{k})=E_\sigma(-\vec{k})$) in the Moire Hubbard model. This symmetry is broken by a nonzero $\mathcal{C}$.

After stacking a second WSe$_2$ layer with a small twist angle $\theta$, the effective Hamiltonian around $\vec{K}_0$ valley is:
\begin{widetext}
\begin{equation}
    \mathcal{H}_{\vec K_0 ,\uparrow}=\int d^2\vec{r}~\Psi^{\dagger}(\vec{r})\left(\begin{array}{cc}h^{t}(\vec{k}-\vec{K})+D & \Delta_{T}(\vec{r}) \\ \Delta_{T}^{\dagger}(\vec{r}) & h^{b}(\vec{k}-\vec{K^\prime})-D\end{array}\right)\Psi(\vec{r}),
\end{equation}
\end{widetext}
where $\Psi(\vec{r})=(\psi_t(\vec{r}),\psi_b(\vec{r}))^\text{T}$ is a two-component spinor with the top and bottom layer components. $D$ represents the effect of the displacement field. The diagonal term $h^{t}(\vec{k}-\vec{K})=h(\mathcal{R}_{\theta / 2} (\vec{k}-\vec{K}))$ is the single layer Hamiltonian for the top layer after a twist angle $\frac{\theta}{2}$. $h^b$ is obtained from $h^t$ by replacing $\theta$ by $-\theta$ and $\vec{K}$ by $\vec{K}^\prime$. And the offdiagonal term $\Delta_{T}$ describes the interlayer tunneling, and is approximated as $\Delta_{T}(\vec{r})=w\left(1+e^{-i \vec{G}_{1} \cdot \vec{r}}+e^{-i \vec{G}_{2} \cdot \vec{r}}\right)$, where $\vec{G}_i$ is the reciprocal lattice vector.  In the strict continuum model with $\mathcal{C}=0$ and $\vec{k}$-independent hybridization the eigenvalues of $h^t(\vec{k})$ are the same as those of $h^b(-\vec{k})$ so that the Moire bands have an effective inversion symmetry. However a nonzero $\mathcal{C}$ combined with the non-zero twist angle means that the eigenvalues of $h^t(\vec{k})$ and $h^b(-\vec{k})$ are not equal except along certain high symmetry lines such as $\Gamma-K$.

\begin{figure}[ht]
  \includegraphics[width=0.9\columnwidth]{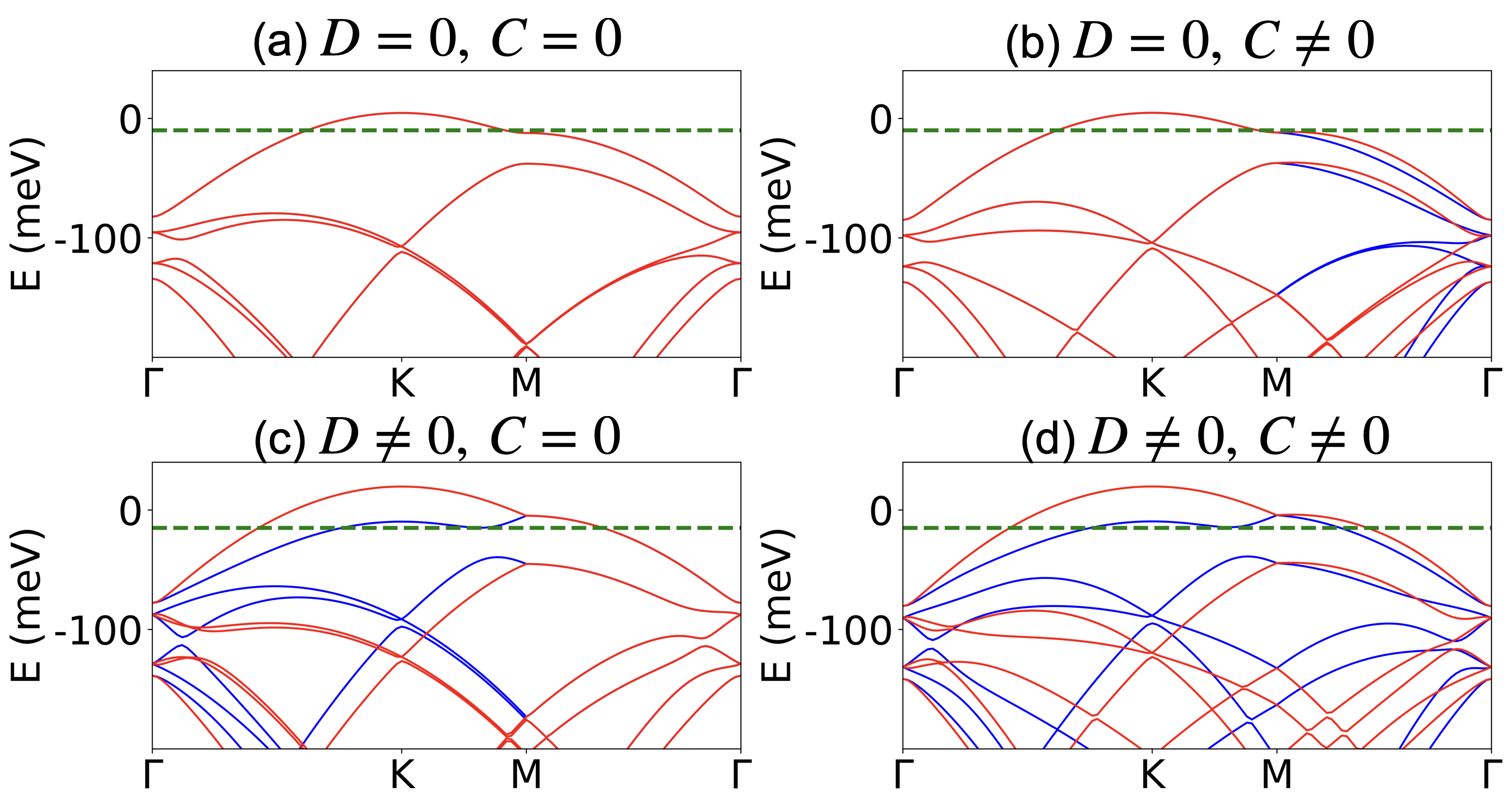}
  \caption{Band structure of the continuum model of \twsetwo~with and without the displacement field $D$ and the high order term  $\mathcal{C}|\vec{k}-\vec{K}_0(^\prime)|^{3} \cos (3 \alpha_{\vec{k}})$. Blue and red solid lines represent $\vec{K}_0$ and $\vec{K}^\prime_0$ valleys. Green dashed lines indicate the energy level of half filling of the topmost valence bands.}
  \label{fig:band_structure}
\end{figure}

After a Fourier transform, with basis $\Psi(\vec{p})=(\psi_t(\vec{p}),~\psi_b(\vec{p}),~\psi_b(\vec{p}+\vec{q}_1),~\psi_b(\vec{p}+\vec{q}_2),...)^\text{T}$, the highest moir\'e band can be viewed as a result of back-folding the monolayer bands into the moir\'e Brillouin zone with spinless hybridization. In Fig.~\ref{fig:band_structure}(a), we plot the band structure for strict continuum  model keeping only the quadratic term $h_0=-(\vec{k}-\vec{K}_0)^{2} / 2 m^\ast$ in the monolayer Hamiltonian (eq.~\ref{eq:Hmonolayer}). If the higher order term $\mathcal{C}|\vec{k}-\vec{K}_0|^{3} \cos (3 \alpha_{\vec{k}})$ is retained (panel (b)), the degeneracy is lifted at general $k$-points; for example  a small splitting along $\vec{\Gamma}$ to $\vec{M}$ in the moir\'e Brillouin zone is evident. These symmetry breaking terms are small for small twist angle because only small deviations of $\vec{k}$ from the single-layer $\vec{K}$ point are relevant. This symmetry breaking term can be described by further neighbor hopping in the moir\'e Hubbard model, with hopping amplitude $\lesssim20\%$  of the first neighbor hopping, and does not change the physics much.  On the other hand (panel (c) and (d)) a non-zero displacement field distinguishes the top and bottom layers and thus strongly splits the degeneracy except along special high symmetry lines such as $\vec{\Gamma}-\vec{M}$ where the symmetry of the monolayer protects the  spin degeneracy.  \cite{Wang:2020us}. 

\section{Mean field approximation }\label{app:meanfield}
The Hubbard model is written as
\begin{align}
	H&=\sum_{\substack{\vec{k},\sigma=\pm}}\epsilon_{\vec{k},\sigma}c^{\dagger}_{\vec{k},\sigma}c_{\vec{k},\sigma}+U \sum_{i} n_{i \uparrow} n_{i\downarrow},
\end{align}
where $\epsilon_{\vec{k},\sigma}=-2|t|\cos (\vec{k} \cdot \vec{a}_{m}+\sigma \phi)$ is the single particle's dispersion of the nearest neighbor tight-binding model. As mentioned in Section \ref{sec:Method}, in the mean field treatment, the interaction is factorized as an approximation shown in Eq.~(\ref{eq:HF}). Here we use 120$^\circ$ spiral order in $x-y$ plane as an example and construct its Hamiltonian.

Assume the averaged spin on site $i$ is $\langle S_i^z\rangle=0$ and $\langle S_i^x\rangle+i\langle S_i^y\rangle= me^{i\vec{Q}\cdot\vec{R}_i}$, where $m$ is the magnetization, and $\vec{Q}=(\pm4\pi/3,0)$ is the wave vector of 120$^\circ$ spiral order. Plus and minus sign indicate different staggered chiralities. We assume the averaged electron density on each site is $\langle n_{i\uparrow} \rangle + \langle n_{i\downarrow}\rangle =n$. Therefore,

\begin{equation}
\langle c_{i,\sigma}^\dagger c_{i,\sigma} \rangle =\left(\begin{array}{cc}
	n/2 & me^{i\vec{Q}\cdot\vec{R}_i} \\ me^{-i\vec{Q}\cdot\vec{R}_i} & n/2\end{array}\right).
\end{equation}

The interaction term is:
\begin{eqnarray}
	V &=& U\sum_ic_{i\uparrow}^\dagger c_{i\uparrow}c_{i\downarrow}^\dagger c_{i\downarrow}-c_{i\uparrow}^\dagger c_{i\downarrow}c_{i\downarrow}^\dagger c_{i\uparrow}\\
	&\approx& U \sum_i  \langle n_{i\uparrow}\rangle n_{i\downarrow}+n_{i\uparrow}\langle n_{i\downarrow}\rangle -\langle n_{i\uparrow}\rangle \langle n_{i\downarrow}\rangle\nonumber\\
	&& -\langle S_i^+\rangle S_i^--\langle S_i^-\rangle S_i^++\langle S_i^+\rangle \langle S_i^-\rangle\nonumber\\
	&=& -mU\sum_{k}(c_{\vec{k}\downarrow}^\dagger c_{\vec{k}-\vec{Q}\uparrow}+c_{\vec{k}\uparrow}^\dagger c_{\vec{k}+\vec{Q}\downarrow})+ UN(n^2/4+m^2).\nonumber
\end{eqnarray}

The Brillouin zone is three-fold, and the basis is chosen as $(c_{\vec{k}-\vec{Q}\uparrow},c_{\vec{k}\uparrow}, c_{\vec{k}+\vec{Q}\uparrow},c_{\vec{k}-\vec{Q}\downarrow},c_{\vec{k}\downarrow},c_{\vec{k}+\vec{Q}\downarrow})^T$. After diagonalizing the Hamiltonian, we find six eigenvalues $e_j$:
\begin{align}\label{eq1}
	e_{1,2}&=\frac{\epsilon_{\vec{k}+\vec{Q},\uparrow}+\epsilon_{\vec{k}-\vec{Q},\downarrow}}{2}\pm\sqrt{(mU)^{2}+(\frac{\epsilon_{\vec{k}+\vec{Q},\uparrow}-\epsilon_{\vec{k}-\vec{Q},\downarrow}}{2})^{2}},\nonumber\\
	e_{3,4}&=\frac{\epsilon_{\vec{k},\uparrow}+\epsilon_{\vec{k}+\vec{Q},\downarrow}}{2}\pm\sqrt{(mU)^{2}+(\frac{\epsilon_{\vec{k},\uparrow}-\epsilon_{\vec{k}+\vec{Q},\downarrow}}{2})^{2}},\nonumber\\
	e_{5,6}&=\frac{\epsilon_{\vec{k}-\vec{Q},\uparrow}+\epsilon_{\vec{k},\downarrow}}{2}\pm\sqrt{(mU)^{2}+(\frac{\epsilon_{\vec{k}-\vec{Q},\uparrow}-\epsilon_{\vec{k},\downarrow}}{2})^{2}}
\end{align} 
And the total energy is written as
\begin{equation}
	E=\sum_{\vec{k}}\sum_j^6e_j(\vec{k})n_{j\vec{k}}+UN(n^2/4+m^2).
\end{equation}

\nocite{*}

\bibliography{apssamp}

\end{document}